\documentclass[11pt,a4paper]{article}
\usepackage{graphicx}
\usepackage{color}
\usepackage{caption}
\usepackage{subcaption}
\usepackage{url}
\usepackage{authblk}
\usepackage{lineno} 
\usepackage{newtxtext,newtxmath}
\usepackage{setspace}
\usepackage{colortbl}
\usepackage{url}
\usepackage[T1]{fontenc}

\title{Detecting directional forces in the evolution of grammar: A case study of the English perfect with intransitives across EEBO, COHA, and Google Books}
\author[1]{Shimpei Okuda}
\author[2]{Michio Hosaka$^{\ast}$}
\author[3]{Kazutoshi Sasahara$^{\dagger}$}
\affil[1]{Graduate School of Informatics, Nagoya University, Furo-cho, Chikusa-ku, Nagoya 464-8601, Japan}
\affil[2]{Department of English Language and Literature, Nihon University, 3-25-40 Sakurajyosui, Setagaya-ku, Tokyo, Japan}
\affil[3]{School of Environment and Society, Tokyo Institute of Technology, 3-3-6 Shibaura Minato-ku, Tokyo 108-0023, Japan}
\date{\small Corresponding authors: \\$^{\ast}$ hosaka@chs.nihon-u.ac.jp\\  $^{\dagger}$ sasahara.k.aa@m.titech.ac.jp}

\affil[]{}
\date{}

\begin{document}
\maketitle

\begin{abstract}
Languages have diverse characteristics that have emerged through evolution. In modern English grammar, the perfect is formed with \textit{have}+PP (past participle), but in earlier English the \textit{be}+PP form also existed. It is widely recognised that the auxiliary verb BE was replaced by HAVE throughout evolution, except for some special cases. However, whether this evolution was caused by natural selection or random drift is still unclear. Here we examined directional forces in the evolution of the English perfect with intransitives by combining three large-scale data sources: EEBO (Early English Books Online), COHA (Corpus of Historical American English), and Google Books. We found that most intransitive verbs exhibited an apparent transition from \textit{be}+PP to \textit{have}+PP, most of which were classified as `selection' by a deep neural network-based model.  These results suggest that the English perfect could have evolved through natural selection rather than random drift, and provide insights into the cultural evolution of grammar.
\end{abstract}

\subsection*{Keywords}
cultural evolution, evolutionary force, human language, natural selection, random drift

\section{Introduction}

\subsection{Cultural evolution of human language}
\label{sec:intro}

There are approximately 7000 languages in the world today, and these languages are unique in terms of their phonology, morphology, and grammar. Such diversity is typically observed in the biological world. Organisms began as prokaryotes, and through the processes of `macro-evolution' (major changes in species occurring over a long period) and `micro-evolution' (small changes in diversity within species), these processes resulted in species of various creatures. In evolutionary linguistics, it is believed that modern languages have been diversified through a similar process~\cite{Dediu2013}; that is, language ability could have emerged in the human species through the macro-evolution, and then it could have led to diverse modern languages in the world through the micro-evolution. The former is called `biological evolution', and the latter is called `cultural evolution'. This paper focuses on the cultural evolution of language, especially grammar.

The study of cultural evolution has dealt with various subjects, including institutions, morality, and religion~\cite{Mesoudi2011}. Unlike the evolution of biological species, most of these phenomena are not visible in traits. It is often said that `language does not fossilise'~\cite{Hauser2002}. The cultural evolution of language has been particularly difficult to study. However, the progress of computational models, analytical methods, and laboratory experiments has led to the development of research on the cultural evolution of language~\cite{Kirby2008, Blythe2012, Kirby2014}. In addition, the appearance of big data and information technology has allowed us to quantify the dynamics of the cultural evolution of language~\cite{Lieberman2007}, as already seen in the evolution of biological species. Notably, Newberry et al.~\cite{Newberry2017} analysed the historical changes in the inflexions of verbs and the auxiliary verb DO in English, demonstrating that these changes are caused not only by natural selection but also by random drift (neutral evolution).

In this paper, we focus on a specific case of the cultural evolution of grammar---auxiliary verb selection in the evolution of the English perfect. Although this phenomenon was previously documented with the analysis of small corpora, we quantify it by using three large-scale data sources, providing insights into the evolutionary dynamics of English grammar. 

\subsection{Auxiliary verb selection and evolution of the perfect}
\label{sec:previous}
It is widely recognised that, in most cases, the auxiliary verb BE was replaced by HAVE throughout the evolution of the English perfect. This phenomenon is referred to as `auxiliary selection'.

First, we review previous studies of the perfect in historical linguistics. In many languages, notably Indo-European languages, the auxiliary verb BE or HAVE is used to construct the perfect~\cite{Aranovich2007-cv, Sorace2000-by, Ackema2017-zi}. In such cases, certain restrictions can be observed in which an auxiliary verb is chosen. In general, HAVE is chosen for transitive verbs, while BE is chosen for intransitive verbs (see (1)). 

\begin{table}[h]
\small
\begin{tabular}{llll}
(1) & a. & Ria \underline{heeft} de schuur geverfd.                & (Dutch)                     \\
    &    & Ria has the shed painted.                   &                             \\
    &    & `Ria has painted the shed.'                 &                             \\
    & b. & Onze nieuwe piano \underline{is} eindelijk gearriveerd. & (Dutch)                     \\
    &    & our new piano is finally arrived            &                             \\
    &    & `Our new piano has finally arrived'         & (Ackema and Sorace 2017:2) \\
\end{tabular}
\end{table}

However, in the case of intransitive verbs, BE is selected for the unaccusative verb, which assigns a theme role to an underlying object (see (2)), while HAVE is selected for the unergative verb which assigns an agent role to its subject (see (3)). This phenomenon is referred to as auxiliary selection.
\begin{table}[h]
\small
\begin{tabular}{llll}
(2) & a. & Ma s{\oe}ur est arriv\'{e}e/*a arriv\'{e} en retard.                   & (French)           \\
    &    & my sister is arrived/has arrived late                      &                    \\
    & b. & Der Zug ist/*hat sp\"{a}t angekommen.                          & (German)           \\
    &    & the train is/has late arrived                              & (Sorace 2004: 256) \\
    &    &                                                            &                    \\
(3) & a. & Les ouvriers ont travaill\'{e}/*sont travaill\'{e}s toute la nuit. & (French)           \\
    &    & the workmen have worked/are worked whole the night         &                    \\
    & b. & Kurt hat/*ist den ganzen Tag gearbeitet.                   & (German)           \\
    &    & Kurt has/is the whole day worked                           & (\textit{ditto})  \\
\end{tabular}
\end{table}

In English, however, there has been a substantial change in the auxiliary verb selection during the process of cultural evolution~\cite{Kyto2011-ht, Huber2019}. In modern English, only HAVE is used as an auxiliary verb (see (4)), but the perfect use of the auxiliary verb BE was often seen (see (5)) from Old English through Late Modern English. It is recognised today that the number of such perfect variations using the verb BE decreased markedly in the 19th century~\cite{Ryden1987}.

\begin{table}[h]
\small
\begin{tabular}{llll}
(4) & a. & John \underline{has/*is} eaten pizza.                                  &                                            \\
    & b. & John \underline{has/*is} worked for an hour.                           &                                            \\
    & c. & John \underline{has/*is} arrived                                       & (McFadden 2007: 675)                       \\
    &    &                                                            &                                            \\
(5) & a. & o{\th}{\th}{\ae}t wintra \underline{bi{\dh}} {\th}usend urnen                              &                                            \\
    &    & until winters(GEN) is thousand run                         &                                            \\
    &    & `until a thousand years have passed'                       & (Phoen 363; Denison 1993: 359)             \\
    & b. & Whanne he escaped \underline{was}                                      & (Chaucer, \textit{CT.Mk.} VII.2735; \textit{ditto}) \\
  & c. & yet Benedicke was such another; and now \underline{is} he become     &                                            \\
    &    &  a man.          & (Shakespeare, Ado III.iv.86; \textit{ditto}) 
\end{tabular}
\end{table}

To analyse the frequency and structural changes of the HAVE and BE perfect, McFadden and Alexiadou~\cite{McFadden2010} used corpora of Old English through Early Modern English (The York-Toronto-Helsinki Parsed Corpus of Old English Prose (YCOE), The Penn-Helsinki Parsed Corpus of Middle English (PPCME2), The Penn-Helsinki Parsed Corpus of Early Modern English (PPCEME), and McFadden~\cite{McFadden2017} used a corpus spanning Early Modern to Modern English (PPCMBE). These corpora are relatively small: YCOE (1.5 M), PPCME2 (1.2 M), PPCEME (1.7 M), and PPCMBE (1.4 M). As a result of these analyses, it was observed that the total frequency of the HAVE perfect increased in Late Middle English (1420--1500), and that of the BE perfect did not decrease until around 1700 and then decreased in the 19th century (1810--1861).\footnote{McFadden and Alexiadou (2010: 422) explains the time gap between the rise of HAVE-perfect and the fall of BE-perfect as follows: ``1350 was not actually the start of the loss of \textit{be}, but the start of an expansion of \textit{have} at the expense of the simple past. Indeed, we have given quantitative evidence that the frequency of the \textit{be} periphrasis was stable throughout Middle English and Early Modern English, that is, up to around 1700.''
}

Previous studies succeeded in identifying some of the factors that influenced the increase of the HAVE perfect, which include linguistic factors such as past counterfactual modality, iterative, durative and atelic meanings, as well as telic eventualities and extralinguistic factors such as chronology and text type~\cite{Kyto2011-ht,Ryden1987,McFadden2010,McFadden2017}. Here, we add quantitative evidence of directional forces for grammar evolution, providing detailed dynamics of each single verb (i.e., evolutionary speed and trajectory) based on computational analysis of large-scale data sources.


With reference to these studies, we aimed to assess whether this change was caused by natural selection or random drift using a large-scale dataset.
We combined three large-scale data sources (as explained in the next section) to detect evolutionary forces from systematically selected verbs in the English perfect using a robust method. The results complement previous findings and provide insights into the cultural evolution of English grammar.

\section{Data and methods}

\subsection{Data}
Table~\ref{tb:corpus} summarises the large-scale data sources used for this study.
The first is Early English Books Online (EEBO), a historical corpus that contains British-related printed matters published between the 15th and 17th centuries~\cite{eebo}. EEBO contains various texts spanning genres, such as literature, history, philosophy, science, politics, law, and economy. Given a query, the EEBO database returns matched sentences with the relevant information, including the corresponding document IDs, pages, and years, from which we can compute the frequency of matched cases (counts).

The second is the Corpus of Historical American English (COHA), comprising American English texts in fiction, popular magazines, newspapers, and non-fiction (books) spanning the 19th to 21st centuries~\cite{coha}.\footnote{COHA was updated in 2021. We used the 2010 version.}  COHA is one of the largest structured historical corpora of American English, and it enables performing queries and providing the matched sentences with their time stamps. Similarly, from this, we can compute the frequency of matched cases (counts).

The third is $N$-gram data from Google Books Ngram Viewer site (English, 2012 version)~\cite{google-ngram-viewer-2012} that is used to incorporate data spanning the 18th to 21st centuries, which EEBO and COHA do not cover.
In this paper, we simply refer to this as Google Books data. 
Using web scraping, we searched and collected both British and American English publications between 1700 and 2000 at the site, as EEBO covers British English 1473--1700 and COHA does American English 1810--2009.

It should be noted that the collected Google Books Ngram Viewer site does not have the Part-of-Speech (POS) tag function; given a simple query, it only provides the relative frequency of words/phrases (NOT raw counts), without matched source texts. 
Thus, we need to adjust the resulting frequencies of Google Books $N$-gram search based on those from the two other corpora, which we call `scaling' and explain later.

\begin{table}[t]
    \begin{center}
        \caption{Three data sources used in this study.}
        \label{tb:corpus}
        \small{
        \begin{tabular}{llr}\hline
            \multicolumn{1}{c}{Corpus} & \multicolumn{1}{c}{Year} & \multicolumn{1}{c}{Size (words)}\\
            \hline
            \hline
            1. Early English Books Online (EEBO) & 1473--1700 & 7.55 M\\
            2. Corpus of Historical American English (COHA) & 1810--2009  & 400 M\\
            3. Google Books & 1700--2000 & 468 B\\\hline
        \end{tabular}
        }
    \end{center}
\end{table}

\subsection{Intransitive verbs}
\label{sec:dict}
As the \textit{be}+PP (past participle) construction is used not only in the perfect but also in the passive, it is difficult for computer programs to automatically distinguish between these in large-scale corpora (EEBO and COHA) and especially in big data without grammatical tags (Google Books). Therefore, in this study, we focused only on `intransitive verbs' as targets to accurately detect the perfect construction of \textit{be}+PP. This is because intransitive verbs are not used in the passive, in principle. Due to this restriction, we cannot cover diverse expressions of the perfect, but instead we can accurately count the occurrences of the perfect in the datasets. To select intransitive verbs, we used the \textit{Longman Dictionary of Contemporary English} (LDOCE) Online~\cite{ldoce}, which is a standard dictionary of the English language.

\subsection{Selection of target verbs}
\label{sec:target_verbs}

To quantify the cultural evolution of the English perfect, we systematically selected intransitive verbs that appeared in all three datasets and had a high frequency within them.  In this study, we used two groups of target verbs: (A) verbs selected based on the frequency and (B) verbs that did not meet the conditions in (A) but were used in prior studies, as shown next. 

\subsubsection*{(A) Verb group selected based on frequency} 
The Corpus of Contemporary American English (COCA; 400M tokens between 1990 and 2019) provides a full list of the most frequent contemporary words ($n=60,000$), and we used this list as a starting point. 

From this list, we filtered all verbs (Step 1) and then filtered intransitive verbs using LDOCE Online mentioned in Section~\ref{sec:dict} (Step 2).
As shown in Table~\ref{tb:verbselect}, limiting the analysis to intransitive verbs decreases the number of target verbs, but it is a necessary procedure. This is because most English verbs exhibit properties of both intransitive and transitive verbs~\cite{Gelderen2018}.

Then, from the filtered intransitive verbs, we selected only those that appeared more than 200 times in each of EEBO, COHA, and Google Books (Step 3). 
This is the strictest condition for our quantification. 
We also examined a mild condition (30 times in each dataset), as described later. 
We set these values based on our pre-investigation results; 1) no target verbs remain when the threshold is larger than 200, and therefore we set it as the upper bound; 2) when the threshold is below 30, the computed frequencies of less popular verbs around the threshold are just unreliable and therefore it was set as the lower bound. 

Finally, we selected only those intransitive verbs with the \textit{be}+PP frequency of 0.5 or greater in our oldest corpus EEBO. This eliminates intransitive verbs for which the \textit{have}+PP form was already dominant in earlier English.
This process aims to ensure that, for all the target verbs, the evolution of the perfect began in the period covered by EEBO, COHA, and Google Books.

\begin{table}[t]
\caption{Steps for the frequency-based verb selection.} 
\label{tb:verbselect}
\small{
\begin{tabular}{lll}
\hline
Step  & Procedure                                                                                                                       & Selected verbs\\
\hline
\hline
1 & Select verbs from a full list of contemporary words in COCA                                                                                      & 5,764              \\
2 & \begin{tabular}[c]{@{}l@{}}Select intransitive verbs based on LDOCE Online\end{tabular} & 719                \\
3 & \begin{tabular}[c]{@{}l@{}}Select verbs from 2 that occurred more than 200 times \\ in the three datasets\end{tabular}      & 46                 \\
4 & Select verbs from 3 with more than 50\% of \textit{be}+PP in EEBO                                                                        & 13 \\
\hline
\end{tabular}
}
\end{table}

After these steps, we obtained 13 intransitive verbs that were in the \textit{be}+PP form in earlier English (EEBO) and sufficiently frequent in the three data sources (1473--2000).

\subsubsection*{(B) Verb group selected based on prior studies} 
In addition to Group A, we investigated additional example verbs. 
Among intransitive verbs that were excluded using the frequency-based method in verb group A, verbs that were already considered in prior studies would be examined separately from those of Group A. Among verbs listed in LDOCE Online with both intransitive and transitive usage, prior studies~\cite{Kyto2011-ht,Ryden1987} indicated that the six verbs were frequently used in the perfect with BE (see Group B in Table~\ref{tb:verbg_a_b}). Therefore, we analysed them separately from Group A and expected to find evolutionary trajectories different from those of Group A, which may give us additional insights into the cultural evolution of grammar. 

\begin{table}[h]
\caption{Target verbs.} 
\label{tb:verbg_a_b}
\small{
\begin{tabular}{ll}
\hline
\multicolumn{1}{c}{Group} & \multicolumn{1}{c}{Selected verbs}    \\
\hline
\hline
A                       & \begin{tabular}[c]{@{}l@{}}arrive, bound, come, creep, degenerate, expire, fall, insist, look, \\ rise, stay, tumble, vanish\end{tabular} \\
\hline
B                       & ascend, become, depart, descend, escape, go \\  
\hline
\end{tabular}
}
\end{table}

\subsection{Detection of evolutionary forces}
\label{sec:detection}
Unidirectional selection, in which new forms replace older forms over generations, often (but not always) produces an S-shaped growth curve when viewed as a change in allele frequency in a population~\cite{Blythe2012a}, which is often accepted as evidence of directional force favouring one variant over others. 
The historical changes in the selection of auxiliary verbs \textit{be}/\textit{have}+PP can also be expressed mathematically within the same framework. 
For detecting evolutionary forces, there are two established methods, the Frequency Increment Test (FIT)~\cite{Newberry2017,Feder2014} and a neural network-based classification (TSC)~\cite{Karsdorp2020}. 
Both methods are based on the Wright-Fisher model~\cite{Ewens2012-rz},  which describes the drift dynamics of two competing types in a population of fixed size $N$ and uses it as the null model.

The FIT is a method to detect selection in time series data, applying it to population genetics experiments and historic DNA samples~\cite{Feder2014}. The FIT rejects `random drift' (the null hypothesis) when the distribution of increments in the frequency of normalised alleles shows a mean value that deviates significantly from zero, which suggests the possibility of evolutionary forces or `selection'. 
However, the FIT has several drawbacks as reported in the literature~\cite{Karjus2020}: it can be sensitive to how the corpus/dataset is  organised  into temporal segments (i.e., binning), and it assumes the normality of data, which the real data often violate.

To resolve these problems, Karsdorp et al. (2020) proposed a deep neural network model, in which the problem of detecting evolutionary forces is formulated as a binary classification task for a given time series data. This is called the neural time series classification (TSC). This model was trained on time series of cultural change simulated by the Wright-Fisher model. They demonstrated that the neural TSC could resolve problems mentioned above associated with the FIT: it has robustness for specific binning methods, and the normality assumption of frequency increments does not play a role in this model. Nevertheless, the neural TSC can consistently and accurately distinguish time series produced by random drift from time series subject to selection pressure. 

Therefore, in this study, we used the neural TSC library~\cite{nnfit} to detect evolutionary forces underlying the transition from \textit{be}+PP to \textit{have}+PP. In addition, for reference, we also used the FIT in our datasets. However, as explained later (and in Supporting Material), our post-hoc power analysis shows that the data size is insufficient to apply the FIT.

\subsection{Data processing}
\begin{figure}[!t]
\center
\includegraphics[width=\textwidth]{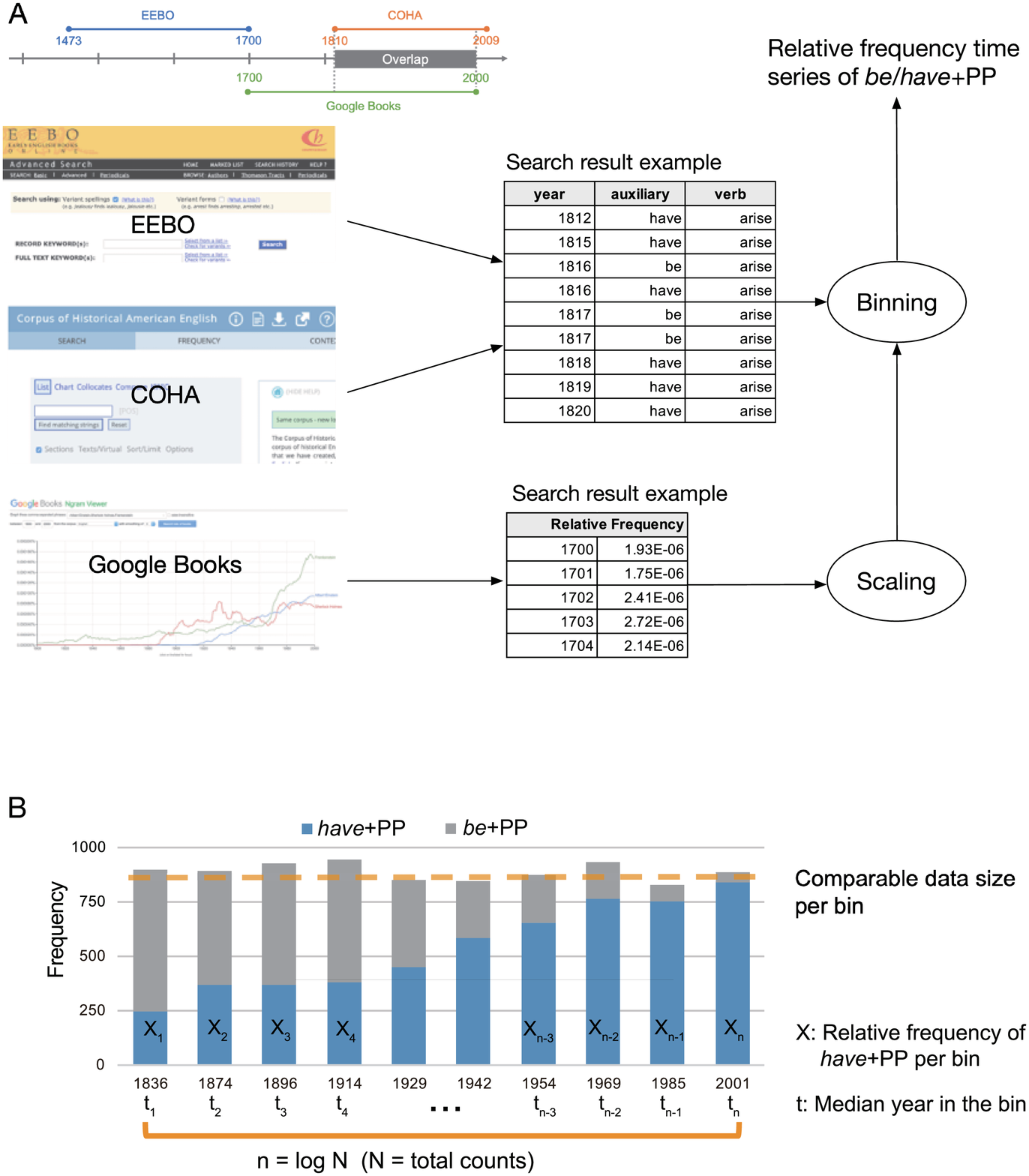}
\caption{Data processing. (A) Construction of the frequency time series for a target verb using three data sources, in which the years represent the time range covered by each corpus in our analysis (B) Binning method. The data size per bin should be comparable.}
\label{fig:method}
\end{figure}
Figure~\ref{fig:method} shows schematic illustrations of our data processing for detecting evolutionary forces. 
As explained in Section~\ref{sec:target_verbs}, we restricted our search queries to the basic form of \textit{be}/\textit{have}+PP in EEBO, COHA, and Google Books to prevent false positives (i.e., \textit{be}+PP for the passive) and accurately compute their frequencies of matches across time for each target verb. Since the spellings of (auxiliary) verbs vary over history, we referred to the formats found in EEBO and formulated lists (Tables S1--3  in Supplementary Material) to perform a comprehensive search of \textit{be}/\textit{have} and the past participle of each verb.

When merging the search results of \textit{be}/\textit{have}e+PP from the three datasets, it should be kept in mind that, for EEBO and COHA, we can retrieve the patterns that match the \textit{be}/\textit{have}+PP construction, along with the year in which the sentence appeared (see Fig.~\ref{fig:method}A upper middle table). By aggregating these data, we can construct a frequency time series given a target verb (raw counts of occurrences per year). As mentioned before, with the Google Books Ngram Viewer, we can only obtain the percentages of matches adjusted by year (relative frequencies of occurrences per year) without the matched patterns. Furthermore, as Google Books is more than three orders of magnitude larger than those of the other two corpora (see Table~\ref{tb:corpus}), we must properly adjust the search results by `scaling'.

Google Books and COHA have an overlapping period between 1810 and 2000. We, therefore, took advantage of this fact to scale the search results from Google Books so that the resulting frequencies roughly matched those from COHA. Specifically, for the years 1810--2000, we focused on all verbs to find the average frequency of \textit{be}/\textit{have}+PP in COHA ($f_C$), and then obtained the average frequency in Google Books ($f_G$) to estimate a scaling constant of $C:=f_C/f_G$. The multiplication of Google Books search results by $C$ was used as data to complement the periods not covered by EEBO and COHA (i.e., 1700--1810). More specifically, we reconstructed the search results for Google Books in as those from EEBO and COHA (like the one in Fig.~\ref{fig:method}A upper middle) based on the scaled frequency.
No scaling was applied to EEBO as the corpus size was comparable to COHA, and there was no overlapping period between these corpora.

Figure~\ref{fig:method}B illustrates the binning for constructing relative frequency time series of \textit{be}/\textit{have}+PP, which used the same setting reported in~\cite{Newberry2017}, where the bin size was set to $\log N$ ($N$ being the total counts). Then, we split the data so that each bin had approximately the same data size, which is a necessary treatment because the neural TSC (and FIT) assumes the frequency changes within the same population (i.e., the Write-Fisher model~\cite{Ewens2012-rz} as the null model). 
The median of the year data in each bin was used as the time for the bin.

The frequency time series of \textit{be}/\textit{have}+PP are shown in Fig. S1 and S2 in Supplementary Material. 
For both Groups A and B, we can confirm that for all verbs, the trend of frequency increase during the period of overlap between Google Books and COHA (1810--2000) is consistent, and the frequency at the border between EEBO and Google Books is mostly consistent. In other words, this indicates that the Google Books results can properly complement the other two corpora through scaling and binning. Although we are aware of little gaps between 1700 and 1750 in the combined time series, this may not considerably affect the results of the neural TSC, because the frequency time series were smoothed by binning and the inflection point seems after these gaps, which was about 1800 according to the literature~\cite{McFadden2010,McFadden2017}.   

\section{Results}
\label{sec:results}

\begin{figure}[t]
    \centering
    \includegraphics[width=\textwidth]{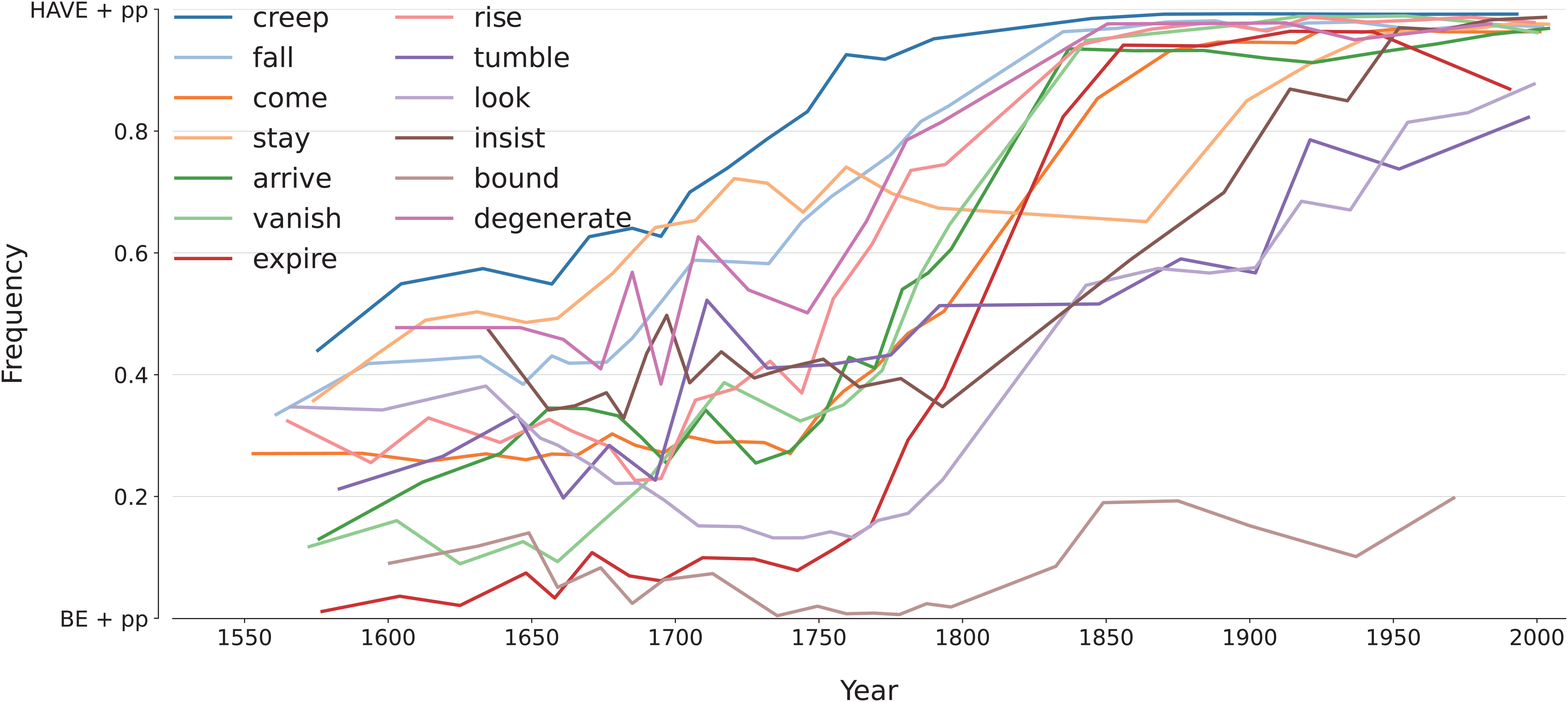}
    \caption{Historical frequency changes of \textit{be}/\textit{have}+PP for the 13 verbs in Group A.}
    \label{fig:freq_a}
\end{figure}

Figure~\ref{fig:freq_a} shows the historical changes in the frequency of \textit{be}/\textit{have}+PP constructions for the 13 verbs in Group A that were selected based on frequency. For all verbs except \textit{bound}, there was a clear increase in frequency from \textit{be}+PP to \textit{have}+PP. While \textit{be}+PP was dominant with most verbs before 1600, there was a sharp increase in the frequency of \textit{have}+PP between 1750 and 1800.
This result clearly shows that the form of \textit{have}+PP became dominant during the cultural evolution of grammar.

\begin{figure}[h]
    \centering
    \includegraphics[width=\textwidth]{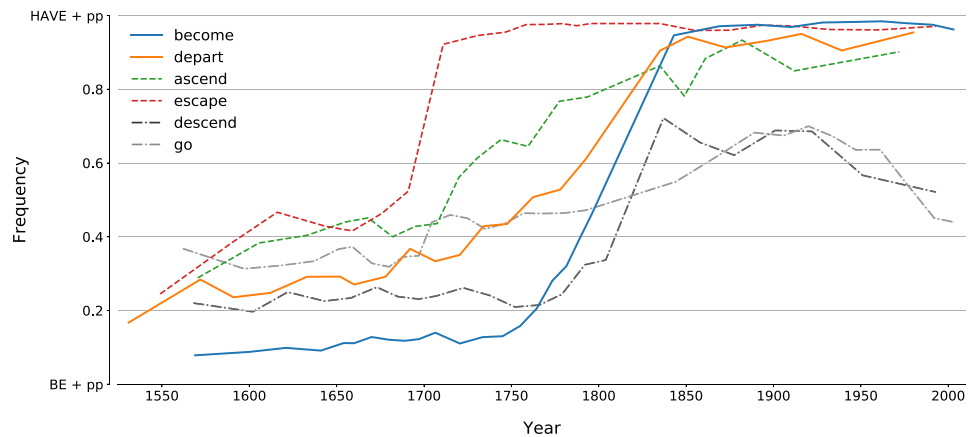}
    \caption{Historical frequency change of \textit{be}/\textit{have}+PP for the six verbs in Group B.}
\label{fig:freq_b}
\end{figure}

Furthermore, we performed the same analysis on the six intransitive verbs from Group B. These verbs were analysed in prior work and excluded from the selection criteria of verb Group A. As shown in Fig.~\ref{fig:freq_b}, similar evolutionary trends can be seen as in Group A; that is, \textit{be}+PP transitions to \textit{have}+PP. 
We also note that the earlier onset (`escape') and the later onset (`descend') are identified in the frequency increase, suggesting individual differences between verbs.

\begin{table}[h]
\caption{Neural TSC result for Group A and B.}
\label{tb:tsc}
\centering
\begin{tabular}{lll}
\hline
Verb       & Group & Probability \\ \hline
arrive     & A     & 1.00       \\
come       & A     & 1.00       \\
creep      & A     & 1.00       \\
degenerate & A     & 1.00       \\
expire     & A     & 1.00       \\
fall       & A     & 1.00       \\
insist     & A     & 1.00       \\
rise       & A     & 1.00       \\
stay       & A     & 1.00       \\
vanish     & A     & 1.00       \\
ascend     & B     & 1.00       \\
become     & B     & 1.00       \\
depart     & B     & 1.00       \\
escape     & B     & 1.00       \\
look       & A     & 1.00       \\
tumble     & A     & 1.00       \\ 
descend    & B     & 0.78       \\ \hline
bound      & A     & 0.07       \\
go         & B     & 0.01       \\ \hline
\end{tabular}
\end{table}

Were the frequency changes of \textit{be}/\textit{have}+PP due to random drift or directional forces at play? The previous studies~\cite{McFadden2010,McFadden2017} did not address this question quantitatively. To examine this, we applied the neural TSC to all the verbs in Group A and B.  
Table~\ref{tb:tsc} summarizes the results, where if the probability is above 0.5, it is deemed as `selection'. 
Of the 19 verbs, 17 verbs were classified as `selection'.
The verbs classified as random drift were `bound' from Group A and `go' from Group B. 
This is visually confirmed from Fig.~\ref{fig:freq_a} and \ref{fig:freq_b}; these verbs did not exhibit an apparent increasing trend or a tipping point in the historical frequency changes of \textit{be}/\textit{have}+PP.
It is likely that these verbs were not classified as `selection' because the increase in \textit{have}+PP was suppressed due to the presence of the \textit{be}+PP passive usage (e.g., `England is bounded on the north by Scotland.' \textit{Is bounded} here is passive.) or the adjective usage of PP (e.g., `His money was gone and his health broken.' \textit{Gone} here can be regarded as an adjective.). 
Note that `bound' is categorised as an intransitive verbs by LDOCE.

For reference, we also conducted the FIT for our datasets. The result is shown in Table S4 in Supplementary material. 
Our post-hoc power analysis revealed that the estimated power $d<0.8$ for all verbs, indicating that the data size was insufficient for this statistical test~\cite{cohen2013statistical}, even though we used the largest data sources. 
Therefore, we did not use the FIT results for discussion.

\section{Discussion}
\subsection*{Evolutionary forces in the cultural evolution of grammar}
In modern English, the perfect takes the form of \textit{have}+PP, and this grammatical rule applies to most verbs. However, in earlier English, there were verbs which formed the perfect with \textit{be}+PP. As mentioned earlier, the prior studies have addressed this phenomenon at the aggregated level; on average, the frequency of the HAVE perfect started increasing in Late Middle English (1420--1500), and that of the BE perfect started decreasing in the 19th century (1810--1861)~\cite{McFadden2010,McFadden2017}. 
Similar trends can be observed in Fig. \ref{fig:freq_a} and \ref{fig:freq_b}. 

In this study, we quantified evolutionary forces by applying a robust neural network-based classification (the neural TSC) to the large-scale data sources: EEBO, COHA, and Google Books with scaling. 
We found that most verbs in Group A and B clearly exhibited an increase in the frequency of \textit{have}+PP, among which 17 verbs were classified as `selection' by the neural TSC. 
Two exceptional cases are explained as the existence of the \textit{be}+PP passive usage or the adjective usage of PP, which can suppress the increase in the frequency of \textit{have}+PP.
They could cause the neural TSC to fail in classification.
Given the fact that most verbs in Group A and B exhibited substantial frequency increases in the form of \textit{have}+PP, it is unlikely that the English perfect evolved through random drift; selection might have played a major role in the evolution of grammar. 
Although previous studies (described in Sec 1.2) identified linguistic and extralinguistic factors accounting for the rise of HAVE and the corresponding decline of BE with intransitives, our analysis allows us to not only reconfirm the tipping point of the BE to HAVE transition at an aggregated level but also provide evolutionary speeds, patterns, and dynamics for each verb. 

The role of random drift in language evolution has been emphasised in a previous study~\cite{Newberry2017}, but our findings are contrasting in that respect. 

We also examined the same experiment using a different setting. 
When lowering the threshold for an expanded selection of target verbs to those appearing at least 30 times in each dataset, comparable results were obtained. 
Most of the target verbs (33/36) were classified as `selection' and the exceptional verbs (`meddle' in addition to `bound' and `go') are all explainable. For example, consider the sentence, `This election was meddled with by the Russians.' Here, \textit{meddled} is passive, although `meddle' is classified as an intransitive verb by LDOCE Online. 
Thus, this result further supports the conclusion that natural selection was the major force for the transition from \textit{be}+PP to \textit{have}+PP.


\subsection*{Limitations}
There are limitations and several potential issues to be considered in our study. 
First, our conclusion is based on the neural TSC that was trained on time series of cultural change simulated by the Wright-Fisher model (as the null model). Thus, if the frequency time series of a verb violated this assumption, the neural TSC could not detect evolutionary forces even if it existed, although the same applies to the other methods, including the FIT. 
Similar to Newberry et al. (2017), the combined datasets of EEBO, COHA, and Google Books include British and American English in a mixture, which might affect the raw counts of matched search results because the pace at which \textit{be}/\textit{have}+PP evolved could be different in British and American English, but the overall tendency holds as we computed the relative frequency of \textit{be}/\textit{have}+PP. 
In addition, there is a `bias' in Google Books resulting from the increasing inclusion of scientific texts~\cite{Pechenick2015}.
Verbs that are commonly used in scientific contexts differ from those that are prevalent in other contexts, and as a result, the relative frequency of \textit{be}+PP / \textit{have}+PP may also vary by genre.
Therefore, it is crucial for future research to acknowledge the significance of genre as a factor in the analysis.

Second, we showed that selection may have played a role in the evolution of the English perfect in many verbs. However, the linguistic reasons for the rise of \textit{have}+PP and the fall of \textit{be}+PP are excluded from our consideration, although linguistic factors, as well as extralinguistic ones, are relevant to the dissipation of \textit{be}+PP, as explained in the previous studies.
Moreover, it is known that throughout its evolution, the perfect has become capable of expressing not only `completion' and `result', but also `experience' and `continuity'. We speculate that the development of such new functions might be related to the shift of auxiliary verbs from BE to HAVE. 
This may be because functional differentiation between BE and HAVE could be adaptive to avoid expressive confusion. 
That is, BE is often used for the passive rather than the perfect, and therefore \textit{be}+PP has always been at risk of being mistaken for the passive. The evolution of the \textit{have}+PP rule reduces such a risk, allowing for the expression of complex meanings. 
This is still at the stage of a hypothesis and needs further investigation. 
Going forward, we should conduct further studies on the relation between the evolution of the perfect and the functional developments within the categories of Aspect and Modality. Moreover, with the same analytical framework, languages other than English should be analysed to investigate the universal dynamics of linguistic diversity resulting from cultural evolution.

\subsection*{Summary}
In summary, our study confirmed the findings of prior work~\cite{Kyto2011-ht,Huber2019,Ryden1987,McFadden2010,McFadden2017} using the larger data sources, and further provided evidence of the possibility of evolutionary forces at work in the transition from \textit{be}+PP to \textit{have}+PP for each verb.
Whether the evolution of the perfect is due to selection or drift in other languages remains an important future research question. 
Our study provides detailed descriptions of processes helpful for conducting such attempts in multiple languages, which is intrinsic to developing the theoretical framework of the cultural evolution of grammar. 

\section*{Data Accessibility}
The data sources used in this study are available:\\ 
EEBO (\url{https://www.english-corpora.org/eebo/}); \\
COHA (\url{https://www.english-corpora.org/coha/}); \\
Google Books (\url{https://books.google.com/ngrams/}).\\ 
The code and process data are available at \url{https://bit.ly/3AVP3F9} and \url{http://bit.ly/3TMQrDy}. (This will be moved to a public repository on publication.)

\section*{Author Contributions}
S.O.: Data curation, formal analysis, investigation, methodology, validation, visualization, writing—review and editing. M.H.: conceptualization, funding acquisition, investigation, writing—review and editing; K.S.: conceptualization, funding acquisition, investigation, writing—original draft, writing—review and editing.
MH and KS designed the study. SO and KS analysed data. All the authors have written and revised the manuscript.

\section*{Competing interests}
We declare that we have no competing interests.

\section*{Ethical statements}
This article does not contain any studies involving human participants.

\section*{Funding}
This work was supported by MEXT/JSPS KAKENHI Grant Numbers \#4903, JP17H06379 and JP17H06383.

\section*{Acknowledgements}
We would like to thank the members of the Evolinguistic project for their fruitful discussions.


\begin{thebibliography}{10}
\expandafter\ifx\csname url\endcsname\relax
  \def\url#1{\texttt{#1}}\fi
\expandafter\ifx\csname urlprefix\endcsname\relax\def\urlprefix{URL }\fi
\providecommand{\bibinfo}[2]{#2}
\providecommand{\eprint}[2][]{\url{#2}}

\bibitem{Dediu2013}
\bibinfo{author}{Dediu, D.} \emph{et~al.}
\newblock \bibinfo{title}{Cultural evolution of language}.
\newblock In \emph{\bibinfo{booktitle}{Cultural evolution: Society, technology,
  language, and religion. Str{\"u}ngmann Forum Reports, vol. 12}},
  \bibinfo{pages}{303--332} (\bibinfo{organization}{MIT Press},
  \bibinfo{year}{2013}).

\bibitem{Mesoudi2011}
\bibinfo{author}{Mesoudi, A.}
\newblock \emph{\bibinfo{title}{Cultural evolution}}
  (\bibinfo{publisher}{University of Chicago Press}, \bibinfo{year}{2011}).

\bibitem{Hauser2002}
\bibinfo{author}{Hauser, M.~D.}, \bibinfo{author}{Chomsky, N.} \&
  \bibinfo{author}{Fitch, W.~T.}
\newblock \bibinfo{title}{{The Faculty of Language: What Is It, Who Has It, and
  How Did It Evolve?}}
\newblock \emph{\bibinfo{journal}{{Science}}} \textbf{\bibinfo{volume}{298}},
  \bibinfo{pages}{1569--1579} (\bibinfo{year}{2002}).

\bibitem{Kirby2008}
\bibinfo{author}{Kirby, S.}, \bibinfo{author}{Cornish, H.} \&
  \bibinfo{author}{Smith, K.}
\newblock \bibinfo{title}{Cumulative cultural evolution in the laboratory: An
  experimental approach to the origins of structure in human language}.
\newblock \emph{\bibinfo{journal}{Proceedings of the National Academy of
  Sciences}} \textbf{\bibinfo{volume}{105}}, \bibinfo{pages}{10681--10686}
  (\bibinfo{year}{2008}).

\bibitem{Blythe2012}
\bibinfo{author}{Blythe, R.~A.}
\newblock \bibinfo{title}{Neutral evolution: A null model for language
  dynamics}.
\newblock \emph{\bibinfo{journal}{Advances in Complex Systems}}
  \textbf{\bibinfo{volume}{15}}, \bibinfo{pages}{1--20} (\bibinfo{year}{2012}).

\bibitem{Kirby2014}
\bibinfo{author}{Kirby, S.}, \bibinfo{author}{Griffiths, T.} \&
  \bibinfo{author}{Smith, K.}
\newblock \bibinfo{title}{Iterated learning and the evolution of language}.
\newblock \emph{\bibinfo{journal}{Current Opinion in Neurobiology}}
  \textbf{\bibinfo{volume}{28}}, \bibinfo{pages}{108--114}
  (\bibinfo{year}{2014}).

\bibitem{Lieberman2007}
\bibinfo{author}{Lieberman, E.}, \bibinfo{author}{Michel, J.-B.},
  \bibinfo{author}{Jackson, J.}, \bibinfo{author}{Tang, T.} \&
  \bibinfo{author}{Nowak, M.~A.}
\newblock \bibinfo{title}{Quantifying the evolutionary dynamics of language}.
\newblock \emph{\bibinfo{journal}{Nature}} \textbf{\bibinfo{volume}{449}},
  \bibinfo{pages}{713--716} (\bibinfo{year}{2007}).

\bibitem{Newberry2017}
\bibinfo{author}{Newberry, M.~G.}, \bibinfo{author}{Ahern, C.~A.},
  \bibinfo{author}{Clark, R.} \& \bibinfo{author}{Plotkin, J.~B.}
\newblock \bibinfo{title}{Detecting evolutionary forces in language change}.
\newblock \emph{\bibinfo{journal}{Nature}} \textbf{\bibinfo{volume}{551}},
  \bibinfo{pages}{223--226} (\bibinfo{year}{2017}).

\bibitem{Aranovich2007-cv}
\bibinfo{author}{Aranovich, R.}
\newblock \emph{\bibinfo{title}{Split Auxiliary Systems: A Cross-linguistic
  Perspective}}.
\newblock Typological studies in language (\bibinfo{publisher}{John Benjamins
  Pub.}, \bibinfo{year}{2007}).

\bibitem{Sorace2000-by}
\bibinfo{author}{Sorace, A.}
\newblock \bibinfo{title}{Gradients in auxiliary selection with intransitive
  verbs}.
\newblock \emph{\bibinfo{journal}{Language}} \textbf{\bibinfo{volume}{76}},
  \bibinfo{pages}{859--890} (\bibinfo{year}{2000}).

\bibitem{Ackema2017-zi}
\bibinfo{author}{Ackema, P.} \& \bibinfo{author}{Sorace, A.}
\newblock \bibinfo{title}{Auxiliary selection}.
\newblock In \emph{\bibinfo{booktitle}{The Wiley Blackwell Companion to Syntax,
  Second Edition}}, \bibinfo{pages}{1--32} (\bibinfo{publisher}{John Wiley \&
  Sons, Ltd}, \bibinfo{year}{2017}).

\bibitem{Kyto2011-ht}
\bibinfo{author}{Kyt{\"o}, M.}
\newblock \bibinfo{title}{Be/have + past participle: The choice of the
  auxiliary with intransitives from {L}ate {M}iddle to {M}odern {E}nglish}.
\newblock In \bibinfo{editor}{Rissanen, M.}, \bibinfo{editor}{Kyt{\"o}, M.} \&
  \bibinfo{editor}{Heikkonen, K.} (eds.) \emph{\bibinfo{booktitle}{Corpus-based
  Studies in Linguistic Variation and Genre Styles}}, \bibinfo{pages}{17--86}
  (\bibinfo{publisher}{De Gruyter Mouton}, \bibinfo{address}{Berlin, Boston},
  \bibinfo{year}{2011}).

\bibitem{Huber2019}
\bibinfo{author}{Huber, J.}
\newblock \bibinfo{title}{{Counterfactuality and aktionsart: Predictors for BE
  vs. HAVE + past participle in Middle English}}.
\newblock In \bibinfo{editor}{Claridge, C.} \& \bibinfo{editor}{B{\"o}s, B.}
  (eds.) \emph{\bibinfo{booktitle}{{Developments in English historical
  morpho-syntax}}}, \bibinfo{pages}{149--173} (\bibinfo{publisher}{John
  Benjamins}, \bibinfo{address}{Amsterdam}, \bibinfo{year}{2019}).

\bibitem{Ryden1987}
\bibinfo{author}{Ryd{\'e}n, M.} \& \bibinfo{author}{Brorstr\"{o}m, S.}
\newblock \emph{\bibinfo{title}{\textit{The} be/have \textit{variation with
  intransitives in {E}nglish. {W}ith special reference to the {L}ate {M}odern
  {P}eriod}}}, vol.~\bibinfo{volume}{70} of \emph{\bibinfo{series}{Stockholm
  studies in English}} (\bibinfo{publisher}{{Almqvist och Wiksell}},
  \bibinfo{year}{1987}).

\bibitem{McFadden2010}
\bibinfo{author}{McFadden, T.} \& \bibinfo{author}{Alexiadou, A.}
\newblock \bibinfo{title}{{P}erfects, {R}esultatives, and {A}uxiliaries in
  {E}arlier {E}nglish}.
\newblock \emph{\bibinfo{journal}{Linguistic Inquiry}}
  \textbf{\bibinfo{volume}{41}}, \bibinfo{pages}{389--425}
  (\bibinfo{year}{2010}).

\bibitem{McFadden2017}
\bibinfo{author}{McFadden, T.}
\newblock \bibinfo{title}{On the disappearance of the {BE} perfect in {L}ate
  {M}odern {E}nglish}.
\newblock \emph{\bibinfo{journal}{Acta Linguistica Hafniensia}}
  \textbf{\bibinfo{volume}{49}}, \bibinfo{pages}{159--175}
  (\bibinfo{year}{2017}).

\bibitem{eebo}
\bibinfo{author}{Lodge, T.}
\newblock \bibinfo{title}{{Early English Books Online}}.
\newblock \bibinfo{howpublished}{\url{https://www.english-corpora.org/eebo/}}
  (\bibinfo{year}{2017}).

\bibitem{coha}
\bibinfo{author}{Davies, M.}
\newblock \bibinfo{title}{{Corpus of Historical American English}}.
\newblock \bibinfo{howpublished}{\url{https://www.english-corpora.org/coha/}}
  (\bibinfo{year}{2010}).

\bibitem{google-ngram-viewer-2012}
\bibinfo{author}{Google}.
\newblock \bibinfo{title}{{Google Ngram Viewer}}.
\newblock \bibinfo{howpublished}{\url{http://books.google.com/ngrams/datasets}}
  (\bibinfo{year}{2012}).

\bibitem{ldoce}
\bibinfo{author}{Education, P.}
\newblock \bibinfo{title}{\textit{Longman Dictionary of Contemporary English}
  {Online}}.
\newblock
  \bibinfo{howpublished}{\url{https://www.ldoceonline.com/jp/dictionary/},
  \url{https://www.ldoceonline.com/jp/dictionary/english-japanese/}}.

\bibitem{Gelderen2018}
\bibinfo{author}{van Gelderen, E.}
\newblock \emph{\bibinfo{title}{\textit{The Diachrony of Verb Meaning:Aspect
  and Argument Structure}}} (\bibinfo{publisher}{Routledge},
  \bibinfo{year}{2018}).

\bibitem{Blythe2012a}
\bibinfo{author}{Blythe, R.~A.} \& \bibinfo{author}{Croft, W.}
\newblock \bibinfo{title}{S-curves and the mechanisms of propagation in
  language change}.
\newblock \emph{\bibinfo{journal}{Language}} \textbf{\bibinfo{volume}{88}},
  \bibinfo{pages}{269--304} (\bibinfo{year}{2012}).

\bibitem{Feder2014}
\bibinfo{author}{Feder, A.~F.}, \bibinfo{author}{Kryazhimskiy, S.} \&
  \bibinfo{author}{Plotkin, J.~B.}
\newblock \bibinfo{title}{{I}dentifying {S}ignatures of {S}election in
  {G}enetic {T}ime {S}eries}.
\newblock \emph{\bibinfo{journal}{Genetics}} \textbf{\bibinfo{volume}{196}},
  \bibinfo{pages}{509--522} (\bibinfo{year}{2014}).

\bibitem{Karsdorp2020}
\bibinfo{author}{Karsdorp, F.}, \bibinfo{author}{Manjavacas, E.},
  \bibinfo{author}{Fonteyn, L.} \& \bibinfo{author}{Kestemont, M.}
\newblock \bibinfo{title}{Classifying evolutionary forces in language change
  using neural networks}.
\newblock \emph{\bibinfo{journal}{Evolutionary Human Sciences}}
  \textbf{\bibinfo{volume}{2}}, \bibinfo{pages}{1--16} (\bibinfo{year}{2020}).

\bibitem{Ewens2012-rz}
\bibinfo{author}{Ewens, W.~J.}
\newblock \emph{\bibinfo{title}{{Mathematical Population Genetics 1:
  Theoretical Introduction}}}.
\newblock Interdisciplinary Applied Mathematics (\bibinfo{publisher}{Springer},
  \bibinfo{address}{New York}, \bibinfo{year}{2012}).

\bibitem{Karjus2020}
\bibinfo{author}{Karjus, A.}, \bibinfo{author}{Blythe, R.~A.},
  \bibinfo{author}{Kirby, S.} \& \bibinfo{author}{Smith, K.}
\newblock \bibinfo{title}{Challenges in detecting evolutionary forces in
  language change using diachronic corpora}.
\newblock \emph{\bibinfo{journal}{Glossa: a journal of general linguistics}}
  \textbf{\bibinfo{volume}{5}}, \bibinfo{pages}{1--25} (\bibinfo{year}{2020}).

\bibitem{nnfit}
\bibinfo{author}{{Folgert Karsdorp}}.
\newblock \bibinfo{title}{\textit{nnfit}}.
\newblock \bibinfo{howpublished}{\url{https://github.com/fbkarsdorp/nnfit}}.

\bibitem{cohen2013statistical}
\bibinfo{author}{Cohen, J.}
\newblock \emph{\bibinfo{title}{Statistical Power Analysis for the Behavioral
  Sciences}} (\bibinfo{publisher}{Taylor \& Francis}, \bibinfo{year}{2013}).

\bibitem{Pechenick2015}
\bibinfo{author}{Pechenick, E.~A.}, \bibinfo{author}{Danforth, C.~M.} \&
  \bibinfo{author}{Dodds, P.~S.}
\newblock \bibinfo{title}{{C}haracterizing the {G}oogle {B}ooks {C}orpus:
  {S}trong {L}imits to {I}nferences of {S}ocio-{C}ultural and {L}inguistic
  {E}volution}.
\newblock \emph{\bibinfo{journal}{{PLOS} {ONE}}} \textbf{\bibinfo{volume}{10}},
  \bibinfo{pages}{1--24} (\bibinfo{year}{2015}).

\end{thebibliography}

\end{document}